\documentclass{article}  
\usepackage{bigsky2009}
\usepackage{graphicx}
\frompage{000} \topage{000}                                              
\def\rightmark{Recent progress in lattice QCD at finite}
\def\leftmark{P. Petreczky}
 
\title{Recent progress in lattice QCD at finite temperature} 
\authors{
{P\'eter Petreczky$^1$  %
}\\[2.812mm]
{\normalsize
\hspace*{-8pt}$^1$ Physics Department, Brookhaven National Laboratory, \\ 
Upton, NY 11973, USA\\[0.2ex] 
}}
 
\abstract{
I review recent progress in finite temperature lattice calculations, 
including the
study of the nature of the deconfinement transition in QCD, 
equation of state, screening of static
quarks and meson spectral functions.
}

 
\begin{document}
 
\maketitle
\setcounter{page}{1}

\section{Introduction}\label{intro}

One expects that at sufficiently high
temperatures and densities the strongly interacting matter undergoes a
transition to a new state, where quarks and gluons are no longer confined in
hadrons, and which is therefore often referred to as a deconfined phase
or quark-gluon plasma.
The main goal of heavy ion experiments is to
create such form of matter and study its properties.
We would like to know at which temperature the transition takes place and what
is the nature of the transition as well the properties of the deconfined
phase, equation of state, static screening lengths, transport properties etc.
Lattice QCD can provide first principle calculation of the transition temperature,
equation of state and static screening lengths (see Ref. \cite{rev} for recent review ). 
In this contribution I am going to review recent progress made in the study of QCD transition
at finite temperature, calculations of equation of state, singlet free energy of static quarks
and meson spectral functions.
 
\section{The finite temperature transition and equation of state}
One of the most interesting question for the lattice
is the question about the nature of the finite temperature transition
and the value of the temperature $T_c$ where it takes place.
For very heavy quarks we have a 1st order deconfining transition.
In the case of QCD with three degenerate flavors of quarks we expect a 1st order
chiral transition for sufficiently small quark masses. 
In other cases there is no true phase transition but just a rapid
crossover.
Lattice simulations of 3 flavor QCD with improved staggered quarks (p4) using
$N_{\tau}=4$ lattices indicate that
the transition is first order only for very small quark masses,
corresponding to pseudo-scalar meson masses of about $60$ MeV
\cite{karschlat03}.
A recent study of the transition using effective models
of QCD  resulted in a similar estimate for the boundary in the quark mass
plane, where the transition is 1st order \cite{szepzs}.
This makes it unlikely that for the interesting case of one heavier strange
quark and two light $u,d$ quarks, corresponding to $140$ MeV pion, the
transition is 1st order. However, calculations with unimproved staggered
quarks suggest that the transition is 1st order for pseudo-scalar
meson mass of about $300$ MeV \cite{norman}.
Thus the effect of the improvement is
significant and we may expect that the improvement of flavor symmetry,
which is broken in the staggered formulation, is very important.
But even when using improved staggered fermions 
it is necessary to do the calculations at several
lattice spacings in order to establish the continuum limit.
Recently,  extensive calculations have been done to clarify the nature
of the transition in the 2+1 flavor QCD for physical quark masses using
$N_{\tau}=4,~6,~8$ and $10$ lattices. These calculations were done using the
so-called $stout$ improved staggered fermion formulations which improves
the flavor symmetry of staggered fermions but not the rotational symmetry,  
The result of this study was
that the transition is not a true phase transition but only a rapid
crossover \cite{nature}. 
New calculations with stout action indicate that
only for quark masses about ten times smaller than the physical quark mass
the transition could be first order \cite{endrodi}.
Even-though there is no true phase transition in  
QCD thermodynamic observables change rapidly in a small
temperature interval and the value of
the transition temperature plays an important role.
The flavor and quark mass dependence of
many thermodynamic quantities is largely determined by the flavor and
quark mass dependence of $T_c$. For example, the pressure normalized by
its ideal gas value for pure gauge theory, 2 flavor, 2+1 flavor and 3 flavor
QCD shows almost universal behavior as function of $T/T_c$ \cite{cargese}.

The chiral condensate $\langle \bar \psi \psi \rangle$ and 
the expectation value of the Polyakov loop $\langle L \rangle$
are order parameters in the limit of vanishing and infinite quark masses respectively. 
However, also for finite values of
the quark masses they show a rapid change in vicinity of the transition temperature. 
Therefore they can be used to locate the transition point. 
The fluctuations of the chiral condensate and Polyakov loop
have a peak at the transition temperature. 
The location of this peak has been used to define the
transition temperature in the calculations with p4 action 
on lattices with temporal extent $N_{\tau}=4$ and $6$ for several
values of the quark mass \cite{us06}. 
The combined continuum and chiral extrapolation then gives the value $T_c=192(7)(4)$MeV.
In this calculations the lattice spacing has been fixed by the Sommer 
parameter $r_0=0.469(7)$fm \cite{gray}. 
The last error in the above value of $T_c$ corresponds to the estimated systematic 
error in the extrapolation.
The transition temperature has been
determined using the so-called $stout$ staggered action and $N_{\tau}=8,~10$ and $12$ lattices.
Here the lattice spacing has been fixed using the kaon decay constant $f_K$ as an input \cite{fodor09}. 
The deconfinement temperature has been found to be 
$170(4)(3)$ MeV determined from the Polyakov loop \cite{fodor09} and $169(3)(3)$MeV 
determined from the strangeness susceptibility. 
This value of the transition temperature is significantly smaller than the value obtained with p4 action.
One reason for this discrepancy could be the fact that $N_{\tau}=4$ and $N_{\tau}=6$ lattices are too
coarse for reliable continuum extrapolation. Calculations on $N_{\tau}=8$ lattices indicate 
a relative shift of the
transition temperature by $5$MeV compared to the $N_{\tau}=6$ results \cite{hot}.
The transition temperature for the chiral transition was found to be $146-157$MeV 
depending on the observable, indicating
that the chiral transition happens before the deconfiment transition 
contrary to the conclusion of Ref. \cite{hot}.
However, it is possible that chiral transition was misidentified in 
Ref. \cite{fodor09} due to the effect of Goldstone
modes below the transition temperature \cite{karsch_sewm08}.

Lattice calculations of equation of state were started some twenty years ago. 
In the case of QCD without dynamical quarks the problem has been solved, 
i.e. the equation of state has been calculated in the continuum limit \cite{boyd96}.
At temperatures of about $4T_c$ the deviation from the ideal gas value 
is only $15\%$ suggesting that quark gluon
plasma at this temperate is weakly interacting. Perturbative expansion of the pressure, 
however, showed very poor
convergence at this temperature \cite{arnold}. 
Only through the use of new re-summed perturbative techniques it was possible
to get agreement with the lattice data \cite{scpt97,braaten,blaizot}. 
To get a reliable calculation of the equation of state on the lattice, 
improved actions have to be used \cite{heller99,karsch00}. 
Equation of state has been calculated using p4 and asqtad
improved staggered fermion actions \cite{milc06,eos_pap} using $N_{\tau}=4$ and $6$ lattices. 
Very recently these calculations have been extended using $N_{\tau}=8$ lattices by the HotQCD collaboration
using both p4 and asqtad actions. 

\begin{figure}
\includegraphics[width=6.5cm]{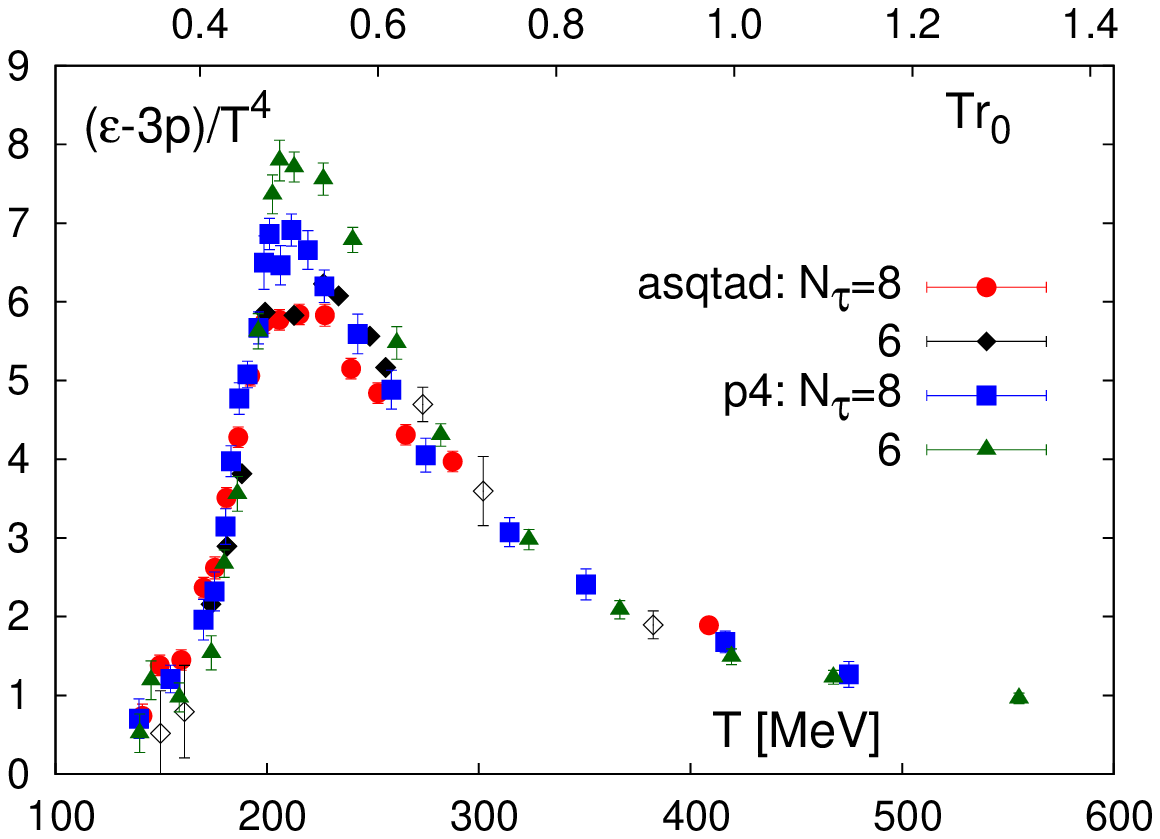}
\includegraphics[width=6.5cm]{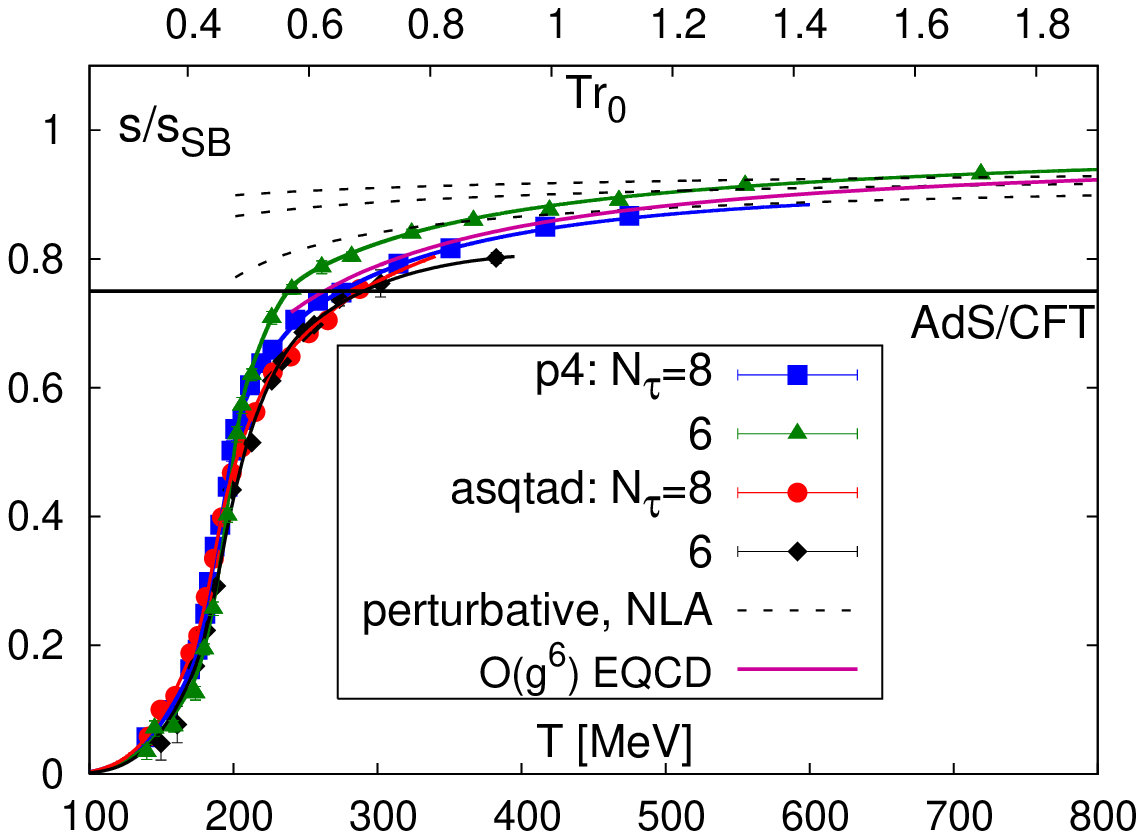}
\caption{ 
The interaction measure calculated (left) and the entropy density divided by the corresponding ideal
gas value (right) calculated with p4 
and asqtad actions \cite{hot}. Comparison with resummed perturbation theory and the weak coupling
expansion is also shown. The solid line is the prediction for strongly coupled gauge theory using
AdS/CFT correspondence \cite{ads/cft}.
}
\label{fig:e-3p}
\end{figure}

In lattice calculations the basic
thermodynamic quantity is the trace of the energy momentum tensor, 
often referred to as the interaction measure $\epsilon-3p$.
This is because it can be expressed in terms of expectation 
values of gauge action and quark condensates (see discussion
in Ref. \cite{eos_pap}). 
All other thermodynamic quantities, pressure, energy density and entropy density $s=(\epsilon+p)$ 
can be obtained from it using integration
\begin{equation}
\frac{p(T)}{T^4} -\frac{p(T_0)}{T_0^4} = \int_{T_0}^T d T' \frac{\epsilon(T')-3 p(T')}{T'^5} 
\end{equation}
The value of $T_0$ is chosen to be sufficiently small so that it corresponds to vanishing pressure
to a fairly good approximation.
In Fig. \ref{fig:e-3p} I show the interaction measure from the new calculations with two actions 
on $N_{\tau}=6$ and $8$ lattices \cite{hot}. At temperatures $T>220$MeV the differences 
between calculations performed on 
$N_{\tau}=6$ and $N_{\tau}=8$ lattices are small, indicating that cutoff effects are under control in this region.
Cutoff effects are seen in the peak region for the p4 action, but not for aqstad action.
In this figure I also show the entropy density which raises rapidly in the temperature region $180-200$ MeV. 
At high temperatures
it is only $10\%$ or less below the ideal gas limit 
in agreement with expectations from improved perturbative calculations \cite{blaizot,rebhan}.
We also compare the results for the entropy density with the weak coupling calculations 
of Ref. \cite{laine06}. The entropy density of  strongly coupled supersymmetric gauge theory is three
quarters of the ideal gas value \cite{ads/cft} and thus is significantly lower than the lattice result.

\section{Color screening in the deconfined phase}
To get further insight into properties of the quark gluon plasma 
one can study  different spatial correlation functions.
One of the most prominent feature of the quark gluon plasma is the presence of chromoelectric (Debye) screening.
The easiest way to study chromoelectric screening is to calculate 
the singlet free energy of static quark anti-quark pair (for recent
reviews on this see Ref. \cite{mehard04,qgp09}),  which is expressed in terms of 
correlation function of temporal Wilson lines in Coulomb gauge
\begin{equation}
\exp(-F_1(r,T)/T)=\frac{1}{N} {\rm Tr} \langle W(r) W^{\dagger}(0) \rangle.
\end{equation}
$L={\rm Tr}  W$ is the Polyakov loop.
Instead of using the Coulomb gauge the singlet free energy can be defined in gauge invariant manner by
inserting a spatial gauge connection between the two Wilson lines. Using such definition the singlet free energy 
has been calculated in SU(2) gauge theory \cite{baza08}.  
It has been found that the singlet free energy calculated this way is close to the result obtained in
Coulomb gauge \cite{baza08}. 
The singlet free energy turned out to be  useful to study quarkonia binding at high temperatures in potential models 
(see e.g. Refs. \cite{digal01,wong,alberico,rapp,mocsy1,mocsy2}).  

\begin{figure}
\includegraphics[width=6.5cm]{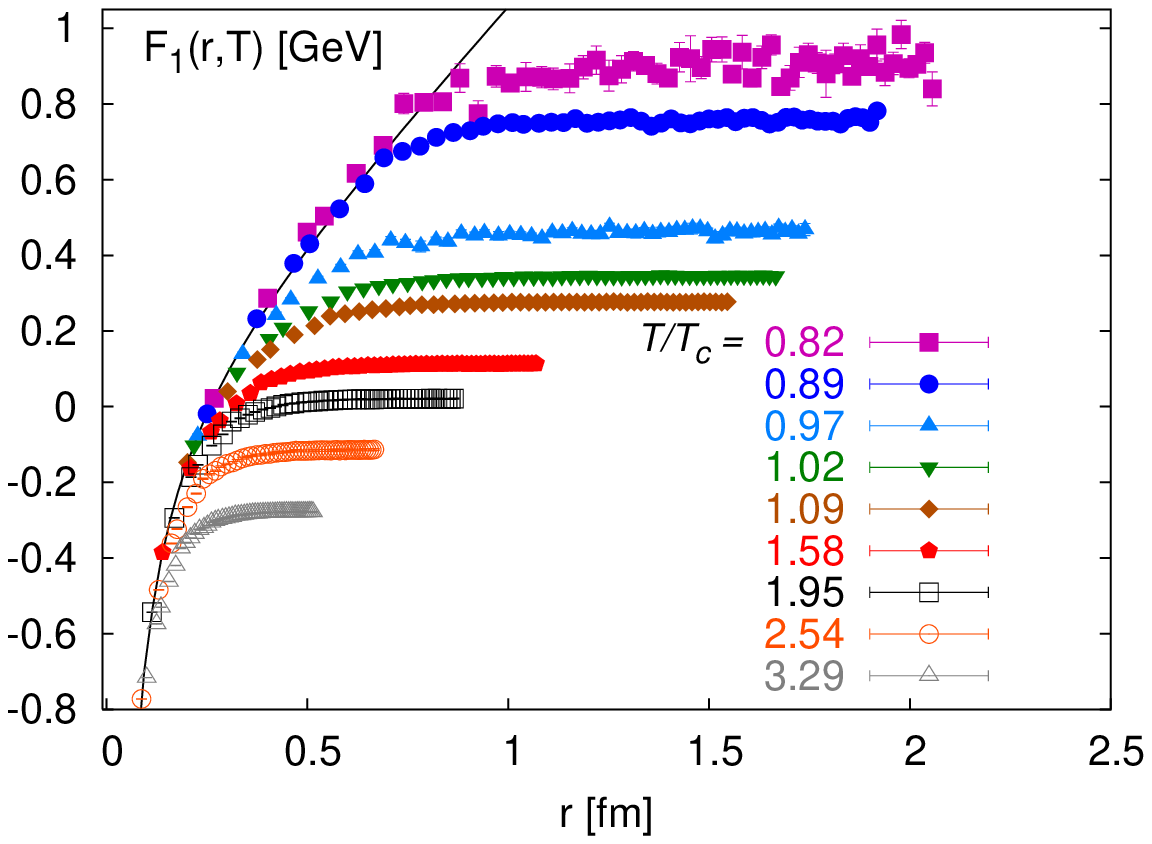}
\includegraphics[width=6.5cm]{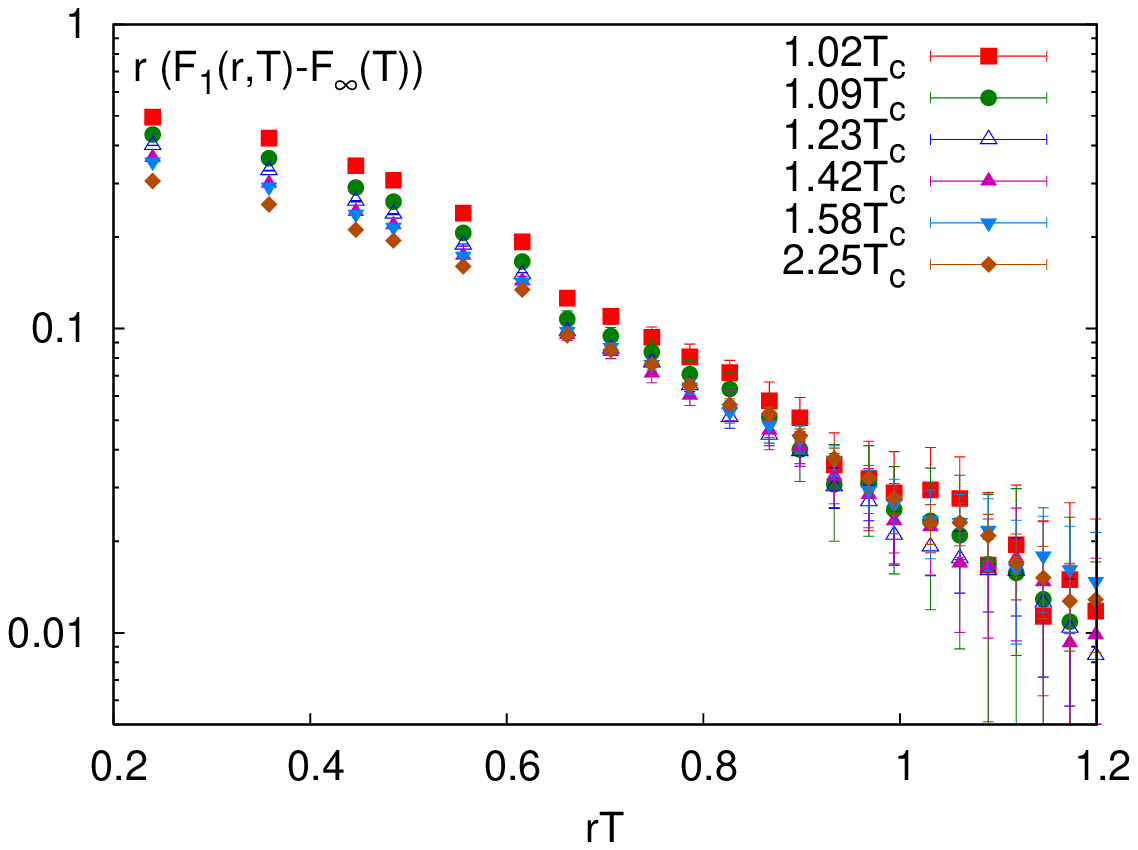}
\caption{The singlet free energy $F_1(r,T)$ calculated in Coulomb gauge on $16^3 \times 4$ lattices (left)
and the combination $F_1(r,T)-F_{\infty}(T)$ as function of $r T$ (right). The solid black line
is the parametrization of the zero temperature potential.}
\label{fig:f1} 
\end{figure}

The  singlet free energy was recently calculated in QCD with one strange quark and two light
quarks with masses corresponding to pion mass of $220$MeV on $16^3 \times 4$ lattices \cite{rbc_f1}. 
The numerical results are shown in Fig. \ref{fig:f1}.
At short distances the singlet free energy is temperature independent and coincides with the 
zero temperature
potential.  In purely gluonic theory the free energy grows linearly with the separation between 
the heavy quark and 
anti-quark in the confined phase. In presence of dynamical quarks the free energy is saturated 
at some finite value 
at distances of about $1$ fm due to string breaking \cite{mehard04,kostya1,okacz05}. 
This is also seen in Fig. \ref{fig:f1}. 
Above the deconfinement temperature the singlet free
energy is exponentially screened at sufficiently large 
distances \cite{okacz04} with the screening mass proportional to
the temperature , i.e.
\begin{equation}
F_1(r,T)=F_{\infty}(T)-\frac{4}{3}\frac{g^2(T)}{4 \pi r} \exp(-m_D(T) r), ~m_D \sim T.
\end{equation}
Therefore in Fig. \ref{fig:f1} we also show the combination $F_1(r,T)-F_{\infty}(T)$ 
as a function of $r T$. As one 
can see from the figure this function shows an exponential fall-off 
at distances $r T>0.8$. The fact that the slope is the same for all temperatures means
that $m_D \sim T$, as expected.

\section{Spectral functions}
Information on hadron properties at finite temperature as well as the transport coefficients 
are encoded in different spectral functions.
In particular the fate of different quarkonium states in the quark gluon plasma 
can studied by calculating the corresponding quarkonium spectral functions 
(for a recent review see Ref. \cite{qgp09}).
On the lattice we can calculate correlation function in Euclidean time. 
This is related to the spectral function via integral
relation
\begin{equation}
G(\tau, T) = \int_0^{\infty} d \omega
\sigma(\omega,T) K(\tau,\omega,T) ,~~
K(\tau,\omega,T) = \frac{\cosh(\omega(\tau-1/2
T))}{\sinh(\omega/2 T)}.
\label{eq.kernel}
\end{equation}
Given the data on the Euclidean meson correlator $G(\tau, T)$ the meson spectral function 
can be calculated
using the Maximum Entropy Method (MEM)  \cite{mem}. For charmonium this was done by 
using correlators calculated on
isotropic lattices \cite{datta02,datta04} as well as  
anisotropic lattices \cite{umeda02,asakawa04,jako07} in quenched approximation.
It has been found that quarkonium correlation function in Euclidean time show only very small temperature
dependence \cite{datta04,jako07}. In other channels, namely the vector, scalar and axial-vector channels 
stronger temperature dependence was found \cite{datta04,jako07}.
The spectral functions in the pseudo-scalar and vector channels reconstructed from MEM show peak structures which may
be interpreted as a ground state peak \cite{umeda02,asakawa04,datta04}. Together with the weak temperature dependence
of the correlation functions this was taken as strong indication that the 1S charmonia ($\eta_c$ and $J/\psi$) survive
in the deconfined phase to temperatures as high as $1.6T_c$ \cite{umeda02,asakawa04,datta04}. A detailed study of
the systematic effects show, however, that the reconstruction of the charmonium spectral function is not reliable
at high temperatures \cite{jako07}, in particular the presence of peaks corresponding to bound states cannot be
reliably established. The only statement that can be made is that the 
spectral function does not show significant changes 
within the errors of the calculations. Recently quarkonium spectral functions have been studied using potential models
and lattice data for the free energy of static quark anti-quark pair \cite{mocsy2}. These calculations show that all
charmonium states are dissolved  at temperatures smaller than $1.2T_c$, but the Euclidean correlators do not show
significant changes and are in fairly good agreement with available lattice data  both 
for charmonium \cite{datta04,jako07} and bottomonium \cite{jako07,dattapanic05}. 
This is due to the fact that even in absence of bound states quarkonium spectral functions
show significant enhancement in the threshold region \cite{mocsy2}.  Therefore previous statements about quarkonia
survival at high temperatures have to be revisited. 

The large enhancement of the quarkonium correlators above deconfinement in the scalar and axial-vector
channel can be understood in terms of the zero mode contribution \cite{mocsy2,umeda07} 
and not due to the dissolution of the $1P$ states as previously thought. 
Similar, though smaller in magnitude, enhancement of quarkonium correlators due to zero mode 
is seen also in the vector channel \cite{jako07}. Here it is related to heavy quark transport \cite{derek,mocsy1}.
Due to the heavy quark mass the Euclidean correlators for heavy quarkonium can be decomposed into
a high and low energy part $G(\tau,T)=G_{\rm low}(\tau,T)+G_{\rm high}(\tau,T)$
The area under the  peak in the spectral functions at zero energy $\omega \simeq 0$ giving the zero mode contribution
to the Euclidean correlator is proportional to some susceptibility, $G^i_{low}(\tau,T) \simeq T \chi^i(T)$.
Therefore it is natural to ask whether it can be described by a quasi-particle model. The generalized
susceptibilities have been calculated in Ref. \cite{aarts05} in the free theory
Replacing the bare quark mass entering in  the expression of the generalized susceptibilities by an 
effective temperature dependent masses one can describe the zero mode contribution very 
well in all channels \cite{me_hq08}.

The spectral function for light mesons as well 
as the spectral function of the energy momentum tensor has been calculated on the lattice
in quenched approximation \cite{karsch02,asakawaqm02,aarts_el,meyer}. However, unlike in the quarkonia case 
the systematic errors in these calculations are not well understood. 

\section{Summary}

Significant progress has been achieved in lattice calculations of thermodynamic quantities 
using improved staggered fermions.
Pressure, energy density and entropy density can 
be reliably calculated at high temperatures when improved actions are
used. Different lattice calculations show that for the physical quark 
masses the transition to the deconfined phase is not
a true phase transition but a crossover. There is some controversy, however, concerning the location of the crossover.
Lattice calculations provide detailed information about screening of static quarks which is important for the fate of
heavy quarkonia in the quark gluon plasma. Some progress has been made in calculating spectral functions on
the lattice, however, much more work is needed in this case. 

\section*{Acknowledgments}
This work was supported by U.S. Department of Energy under
Contract No. DE-AC02-98CH10886. I am grateful to Anton Rebhan for sending his results for entropy density
in the resummed perturbative calculations.

\vfill\eject
\end{document}